\documentclass[aps,prb,twocolumn,footinbib,groupedaddress,longbibliography]{revtex4-1}
\usepackage{graphicx}
\usepackage{epstopdf}
\pdfoutput=1
\usepackage[colorlinks=true, letterpaper=true, pdfstartview=FitV, linkcolor=blue, citecolor=blue, urlcolor=blue]{hyperref}
\usepackage{graphicx}
\usepackage{subfigure}
\usepackage{dcolumn}
\usepackage{bm}
\usepackage{times}
\usepackage{amsmath}
\usepackage{amssymb}
\usepackage{float}
\usepackage{color}
\usepackage{multirow}
\usepackage{url,hyperref}
\usepackage[normalem]{ulem}

\begin{document}

\draft

\title{Electric field controlled valley-polarized photocurrent switch based on the circular bulk photovoltaic effect}

\author{Yaqing Yang,$^{1,5}$ Xiaoyu Cheng,$^{1,5}$ Liantuan Xiao,$^{1,5}$ Suotang Jia,$^{1,5}$ Jun Chen,$^{2,5,*}$ Lei Zhang,$^{1,5,\dagger}$ and Jian Wang$^{3,4,\ddag}$}
\address{$^1$State Key Laboratory of Quantum Optics and Quantum Optics Devices, Institute of Laser Spectroscopy, Shanxi University, Taiyuan 030006 China\\
$^2$State Key Laboratory of Quantum Optics and Quantum Optics Devices, Institute of Theoretical Physics, Shanxi University, Taiyuan 030006, China\\
$^3$College of Physics and Optoelectronic Engineering, Shenzhen University, Shenzhen 518060, China\\
$^4$Department of Physics, The University of Hong Kong, Pokfulam Road, Hong Kong, China\\
$^5$Collaborative Innovation Center of Extreme Optics, Shanxi University, Taiyuan 030006, China}

\begin{abstract}
Efficient electric manipulation of valley degrees of freedom is critical and challenging for the advancement of valley-based information science and technology. We put forth an electrical scheme, based on a two-band Dirac model, that can switch the fully valley-polarized photocurrent between $K$ and $K'$ valleys using the circular bulk electro-photovoltaic effect. This is accomplished by applying an out-of-plane electric field to the two-dimensional valley materials, which enables continuous tuning of the Berry curvature and its sign flip. We found that the switch of the fully valley-polarized photocurrent is directly tied to the sign change of Berry curvature, which accompanies a topological phase transition, for instance, the quantum spin Hall effect and the quantum valley Hall effect. This scheme has been confirmed in monolayer BiAsI$_2$ and germanene through first-principles calculations. Our paper offers a promising strategy for the development of a volatile valley-addressable memory device and could inspire further research in this area.
\end{abstract}

\maketitle

\emph{Introduction}-Recently, significant research interest has emerged in the development of electronic devices that utilize factors other than an electron's charge degrees of freedom \cite{cao2012,isberg2013,hirohata2014,urbaszek2015,yu2017,schaibley2016}. Along with spintronics \cite{candini2011,otani2017,gregersen2017,ahn20202}, the valley degrees of freedom in solid materials, such as graphene \cite{xiao2007,rycerz2007,sui2015} and transition metal dichalcogenides \cite{xiao2012,mak2014,lee2016,zhu2014}, can be harnessed for information encoding, storage, and transportation \cite{xu2016,yu2020,Wang2023}. To use valley degrees of freedom as information carriers, it is necessary to generate and efficiently control the valley polarization \cite{li2014,zhao2017,zeng2012,mak2012}. Electric field tuning is a low-power and efficient approach applied to certain two-dimensional valley materials to adjust valley polarization \cite{ezawa2012,shimazaki2015,zhang2021e,ye2016e,li2018e,liang2023,zhou2021s,li2020e,khan2021,yu2015e,Li2023,yuan2014,liu2018e,Luo2017,Sharma2023}. However, the direct use of electric-induced valley switching as a volatile valley-addressable memory device remains unexplored.

When two-dimensional valley materials are exposed to circularly polarized light, electrons in $K/K^{\prime}$ valleys can be selectively excited depending on the light polarization. This process produces a fully valley-polarized photocurrent \cite{lei2014,sipe2000,xu2018E}. Recently, the out-of-plane electric field has been presented to control the direction and magnitude of the charge photocurrent due to the electrically switchable Berry curvature dipole \cite{xu2018E}. As such, it is worth exploring if the fully valley-polarized photocurrent can be electrically toggled between $K$ and $K^{\prime}$ valleys by adjusting the quantum geometrical quantity, for instance, Berry curvature. It is our work to fill this gap.

In this paper, we propose a scheme to effectively control and switch the fully valley-polarized photocurrent through an out-of-plane electric field. Based on a two-band Dirac model, we analytically show that the fully valley-polarized photocurrent can be switched between $K$ and $K'$ valleys through the circular bulk electro-photovoltaic effect (CBEPV). The fundamental physical mechanism involves changing the sign of Berry curvature in two-dimensional valley materials by applying an out-of-plane electric field. It is found that this change accompanies a topological phase transition between the quantum spin Hall effect (QSHE) and the quantum valley Hall effect (QVHE), or a transition between different QVHE states. The change in the sign of the Berry curvature alters both the valley-dependent optical selection rule and the CBEPV-induced valley-dependent photocurrent, which leads to the switching of the fully valley-polarized photocurrent. Based on atomic first-principles calculations, we present that the fully valley-polarized photocurrent in monolayer BiAsI$_2$ and germanene can be switched between $K$ and $K'$ valleys by tuning the out-of-plane electric field. Our paper demonstrates the significant potential of two-dimensional valley materials for use in volatile valley-addressable memory devices.

\begin{figure}
\centering
\includegraphics[scale=0.75]{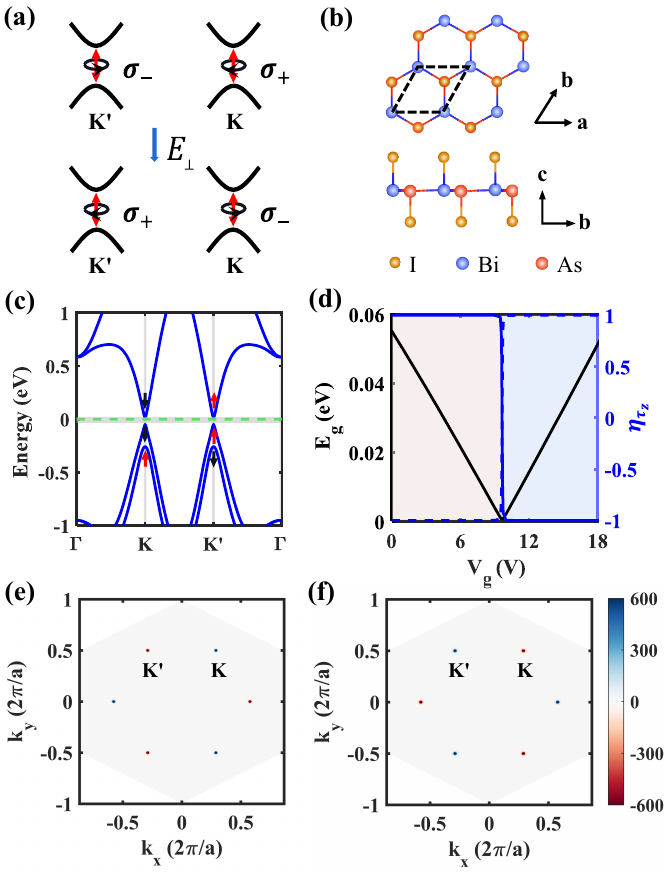}
\caption{(a) Schematic of proposed valley-dependent optical selection rules switching at $K$ and $K'$ by applying an out-of-plane electric field $E_{\perp}$. (b) Crystal structure and (c) band structure of monolayer BiAsI$_2$. The horizontal green dashed line denotes the Fermi energy. (d) The band gap $E_g$ and the degree of circular polarization $\eta_{\tau_z}$ at $K$ and $K^{\prime}$ valleys vs the applied perpendicular gate voltage $V_g$. The solid and dashed blue lines represent $K$ and $K^{\prime}$ valleys, respectively. (e, f) The Berry curvatures of the valence band for monolayer BiAsI$_2$ in the Brillouin zone when $V_g=0,18$ V, respectively.}
\label{Fig1}
\end{figure}

\emph{Berry curvature analysis}-We start from the interband optical transition at a $\bf k$ point in the two-band model. The degree of circular polarization $\eta_{\tau_z}({\bf k})$ can be expressed as \cite{Yao2008}
\begin{equation}
\eta_{\tau_z}({\bf k})=\frac{e}{2\hbar}\frac{\Omega^{v}_{\tau_z}({\bf k})}{\mu^*_B}E_{{\bf k},21}\label{CP}
\end{equation}
where $\Omega^{v}_{\tau_z}({\bf k})$ is the out-of-plane Berry curvature of the valance band. $\tau_z=\pm 1$ denotes $K$ and $K'$ valleys, respectively. $E_{{\bf k},21}=E_{{\bf k},2}-E_{{\bf k},1}$ is the energy difference between two bands at the ${\bf k}$ point, $1$ and $2$ correspond to valence and conduction bands, and $\mu^*_B$ is the effective Bohr magneton. The $\eta_{\tau_z}$ = 1 and -1 values correspond to the absorption of only left ($\sigma_+$) and right ($\sigma_-$) circularly polarized light, respectively. This suggests that if the Berry curvature's sign for a specific valley is reversed by applying the out-of-plane electric field, while the sign of $E_{{\bf k},21}$ is maintained, the corresponding valley-dependent optical selection rules at $K$ and $K'$ can be switched, which is depicted in Fig. \ref{Fig1} (a).

Circularly polarized light is known to produce an injection current in materials when the time reversal symmetry is maintained while the inversion symmetry is broken. The circular photovoltaic/photogalvanic effect (CPGE) induced photocurrent is given by $dJ/dt=dJ^{\sigma_+}/dt-dJ^{\sigma_-}/dt=\beta^{a}_{bc}E_b(\omega)E_c(-\omega)$ \cite{de2017,sipe2000,xu2021p}, where $\beta^{a}_{bc}$ is the CPGE tensor and $E_{b,c}$ is the optical electric field. For a two-band model containing the valence band and the conduction band, the valley-dependent CPGE tensor can be expressed as \cite{sipe2000,de2017}
\begin{equation}
\begin{aligned}
\label{CPGE}
\beta^{a,\tau_z}_{bc}(\omega)&=\sum_{{\bf k}\in \tau_z}\beta^{a}_{bc}(\bf k)\\&=\sum_{{\bf k}\in \tau_z} \frac{i\pi e^3}{\hbar^2V} \partial _{k_a} E_{{\bf k},12}\Omega^{v}_{\tau_z}({\bf k})\delta(\hbar\omega-E_{{\bf k},21})
\end{aligned}
\end{equation}
where $a, b$ and $c$ indicate Cartesian indices, $a$ indicates the direction of the current, and $b$ and $c$ are the polarization directions of the optical electric field. At a given frequency $\omega$, the delta function selects the surface in {\bf k} space where $E_{{\bf k},21}=\hbar\omega$. Note that the CPGE induced valley-dependent photocurrent changes its sign if the Berry curvature changes its sign. If $\eta_{K/K'}=\pm 1$ with no applied out-of-plane electric field, the CPGE tensor $\beta^{a,K/K'}_{bc}$ originates purely from the left/right circularly polarized light contribution. When the out-of-plane electric field is applied, $\eta_{K/K'}$ becomes $\mp 1$ and then $\beta^{a,K/K'}_{bc}$ is purely from the right/left circularly polarized light contribution. Consequently, the valley-dependent CPGE tensor $\beta^{a,K/K'}_{bc}$ changes its sign. This suggests that the fully valley-polarized current can be switched between $K$ and $K'$ valleys when a specific handed polarized light is shining. Therefore, it is essential to find a kind of system that can use out-of-plane electricity to adjust the sign of the Berry curvature.

To achieve this, we consider the Dirac Hamiltonian \cite{liu2011Q,liu2011,zhou2021,Bampoulis2023}
\begin{equation}
H=\hbar v_f(k_x\hat{\sigma}_x+\hat{\tau}_zk_y\hat{\sigma}_y) + \lambda_{SO}\hat{s}_z\hat{\tau}_z\hat{\sigma}_z+U\hat{\sigma}_z,\label{Ham}
\end{equation}
where $v_f$ is the Fermi velocity. $\hat{\sigma}$ is the Pauli matrix denoting orbital. $\hat{s}_z$ and $\hat{\tau}_z$ represent the spin and valley operator, respectively. $\lambda_{SO}$ is the intrinsic spin orbital coupling (SOC) strength. $U$ stands for the staggered potential, which can be controlled by the out-of-plane electric field $E_{\perp}$. By diagonalizing the Hamiltonian in Eq. (\ref{Ham}), we find that the energy spectrum is
\begin{equation}
E=\pm \sqrt{(U+\lambda_{SO}s_z \tau_z)^2 +\hbar^2 k^2 v_f^2}.\label{es}
\end{equation}
Here, $s_z=\pm 1$ represents the spin up and down components, respectively.

The spin-dependent out-of-plane Berry curvature of the valance bands in different valleys is given by
\begin{equation}
\Omega^{s_z}_{\tau_z}({\bf k}) = -\frac{\hbar^2 v_f^2 \tau_z (U+\lambda_{SO}s_z \tau_z)} {{((U+\lambda_{SO}s_z \tau_z)^2 +\hbar^2 k^2 v_f^2})^{3/2}}.\label{bc}
\end{equation}
From Eq. (\ref{bc}), we can know that the sign of Berry curvature $\mathrm{sgn}(\Omega^{\uparrow}_{K},\Omega^{\downarrow}_{K},\Omega^{\uparrow}_{K'},\Omega^{\downarrow}_{K'})=(-,+,-,+)$ when $|U|<\lambda_{SO}$. The topological charge $C^{s_z}_{\tau_z}$ can be obtained by integrating the Berry curvatures. In this situation, the system resides in the QSHE phase with $(C^{\uparrow}_{K},C^{\downarrow}_{K},C^{\uparrow}_{K'},C^{\downarrow}_{K'})=(-0.5,0.5,-0.5,0.5)$ and spin Chern number $-1$. When $|U|>\lambda_{SO}$, the sign of Berry curvature becomes $\mathrm{sgn}(\Omega^{\uparrow}_{K},\Omega^{\downarrow}_{K},\Omega^{\uparrow}_{K'},\Omega^{\downarrow}_{K'})=\mathrm{sgn}(U)(-,-,+,+)$. Correspondingly, the system is in the QVHE phase with $(C^{\uparrow}_{K},C^{\downarrow}_{K},C^{\uparrow}_{K'},C^{\downarrow}_{K'})=\mathrm{sgn}(U)(-0.5,-0.5,0.5,0.5)$.

We can reverse the sign of the Berry curvature by tuning $U$ via the external electric field. According to Eqs. (\ref{es}) and (\ref{bc}), the top of the valence band, when $0<U<\lambda_{SO}$, is spin down/up in $K/K'$ valleys with sign of Berry curvature $\mathrm{sgn}(\Omega^{\downarrow}_{K},\Omega^{\uparrow}_{K'})=(+,-)$ . As $U$ increases beyond $\lambda_{SO}$, the sign of Berry curvatures in both valleys reverses, i.e., $\mathrm{sgn}(\Omega^{\downarrow}_{K},\Omega^{\uparrow}_{K'})=(-,+)$. Similarly, when $0<-U<\lambda_{SO}$, the top of the valence band is spin up/down in $K/K'$ valleys with sign of Berry curvature $\mathrm{sgn}(\Omega^{\uparrow}_{K},\Omega^{\downarrow}_{K'})=(-,+)$. As $-U$ increases beyond $\lambda_{SO}$, the sign of Berry curvatures in both valleys also reverses, i.e., $\mathrm{sgn}(\Omega^{\uparrow}_{K},\Omega^{\downarrow}_{K'})=(+,-)$. Moreover, one can directly tune $U$ between the $-U>\lambda_{SO}$ and $U>\lambda_{SO}$ regimes. As a result, the sign of Berry curvature can be switched between $\mathrm{sgn}(\Omega^{\uparrow}_{K},\Omega^{\downarrow}_{K'})=(+,-)$ and $\mathrm{sgn}(\Omega^{\downarrow}_{K},\Omega^{\uparrow}_{K'})=(-,+)$ with spin switching in the specific valley.


\begin{figure}
\centering
\includegraphics[scale=0.59]{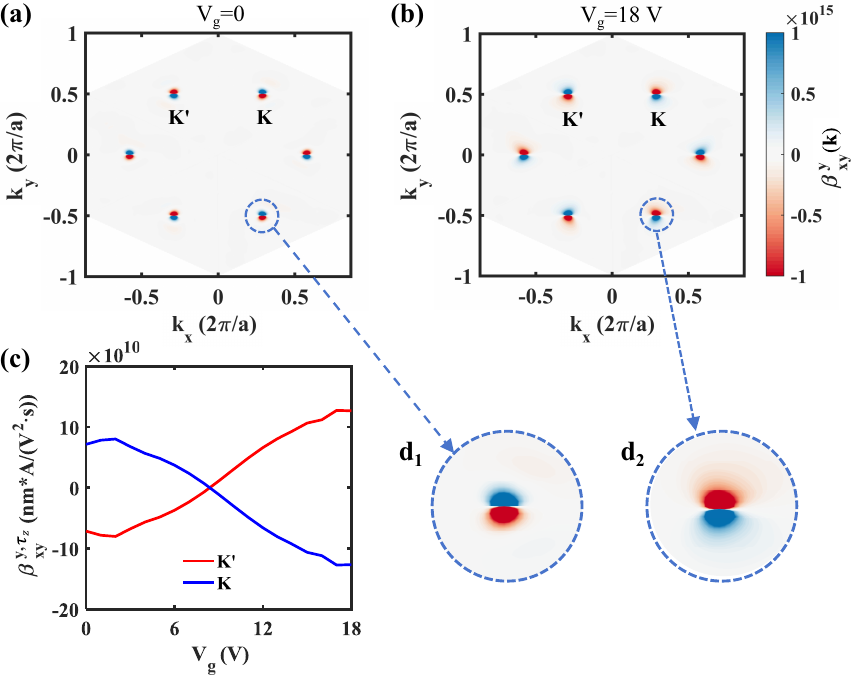}
\caption{(a,b) The $\bf k$-specified contribution to the CPGE tensor $\beta^{y}_{xy}({\bf k})$ of monolayer BiAsI$_2$ in the Brillouin zone when $V_g=0, 18$ V, respectively. (c) The valley-dependent CPGE tensor $\beta^{y,\tau_z}_{xy}$ vs the gate voltage $V_g$. Here, the incident photon energy is fixed at $\hbar\omega=0.06$ eV. ($d_1$,$d_2$) Zoom in on the distribution of $\beta^{y}_{xy}({\bf k})$ near the K valley when $V_g=0, 18$ V, respectively.}
\label{Fig2}
\end{figure}

\emph{Fully valley-polarized photocurrent switch}-After understanding the proposed underlying mechanism of valley switching, we illustrate the switch of fully valley-polarized photocurrent through an out-of-plane electric field in monolayer BiAsI$_2$ first. The crystal structure of monolayer BiAsI$_2$, which has a hexagonal lattice with broken inversion symmetry, is depicted in Fig. \ref{Fig1} (b). Fig. \ref{Fig1} (c) shows the electronic properties of monolayer BiAsI$_2$ and its low energy physics can be well described by Eq. (\ref{Ham})\cite{zhou2021}. Based on the first-principles calculations,\cite{supple}(see also references \cite{kleinman1982,Kresse1993} therein) the monolayer BiAsI$_2$ is in a QSHE state with a direct bandgap of approximately $58$ meV when SOC is considered. The maximum of the valence band and the minimum of the conduction band are located at the $K/K^{\prime}$ valley with spin down/up components. When an external out-of-plane electric field is applied, the bandgap of monolayer BiAsI$_2$ gradually decreases and closes as the electric field increases. As the field continues to increase, the bandgap reopens, and monolayer BiAsI$_2$ undergoes a topological phase transition (TPT) from QSHE to QVHE, as illustrated in Fig. \ref{Fig1} (d) and Supplemental Materials Fig. S1\cite{supple}. Figs. \ref{Fig1} (e) and (f) display the Berry curvature distribution for the valence band in the first Brillouin zone before and after the TPT, respectively. It can be observed that the Berry curvature in the $K/K^{\prime}$ valley changes sign when an external out-of-plane gate voltage is applied. Correspondingly, $\eta_{\tau_z}$ at $K/K'$ switches from $\pm 1$ to $\mp1$, as shown in Fig. \ref{Fig1}(d).

\begin{figure*}
\centering
\includegraphics[scale=0.63]{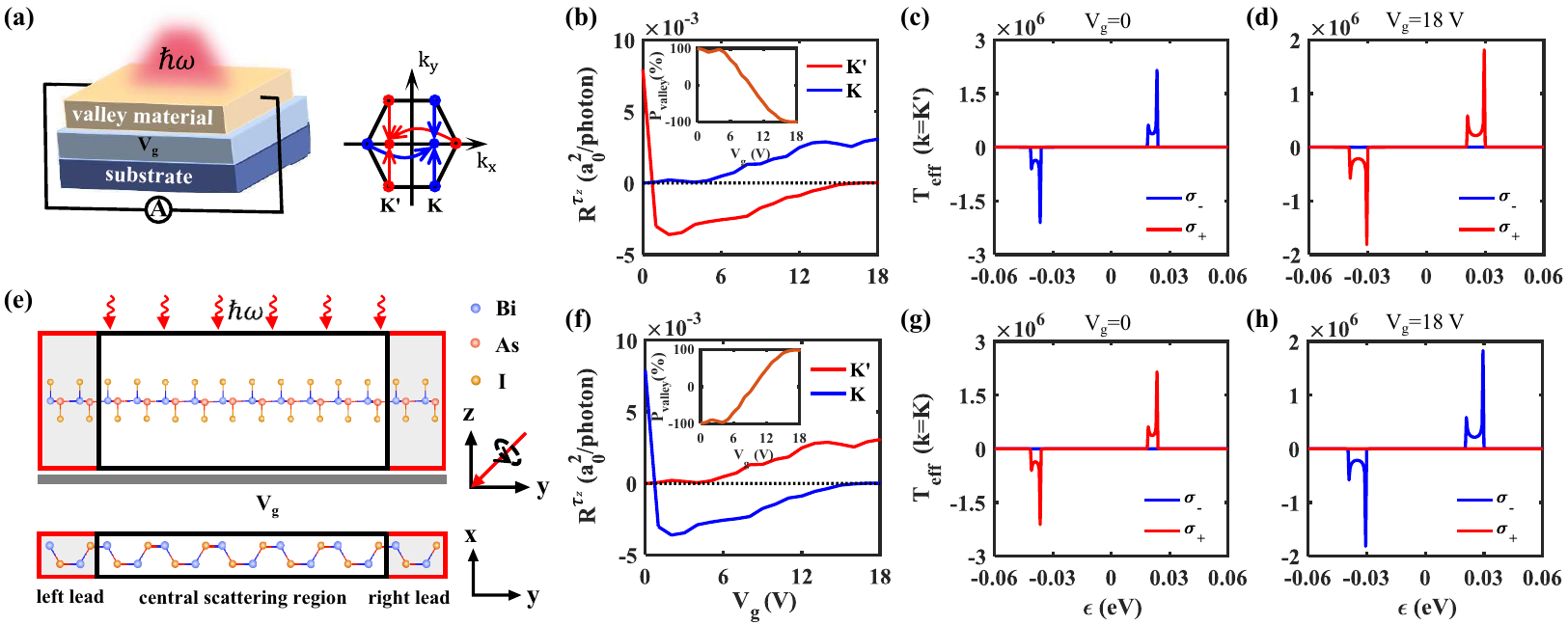}
\caption{(a) The proposed opto-valleytronic device based on two-dimensional valley materials and the corresponding folded Brillouin zone. (e) Schematic plot of the two-probe device constructed by monolayer BiAsI$_2$. The blue, red and yellow spheres represent Bi, As, and I atoms, respectively. (b, f) The valley-polarized photoresponse $R^{\tau_z}$ vs the gate voltage $V_g$ under the vertically incident (b) $\sigma_-$, (f) $\sigma_+$ circularly polarized light, respectively. Inserts show the corresponding valley polarization vs $V_g$. (c, g) The effective transmission $T_{eff}$ vs $\epsilon$ at $K^{\prime}$ and $K$ points when $V_g=0$ V. (d, h) The effective transmission $T_{eff}$ vs $\epsilon$ at $K^{\prime}$ and $K$ points when $V_g=18$ V. Here, the incident photon energy is fixed at $\hbar \omega=0.06$ eV.}
\label{Fig3}
\end{figure*}

Next, we calculate the $\bf k$-specified CPGE tensor $\beta^y_{xy}({\bf k})$ of monolayer BiAsI$_2$ as shown in Fig. \ref{Fig2} (a). We find that the CPGE tensor $\beta^y_{xy}({\bf k})$ at $K$ and $K^{\prime}$ valleys has a ``dipole" distribution with opposite signs. From Eq. (\ref{CPGE}), we learn that $\beta^y_{xy}({\bf k})$ also depends on $\partial E$/$\partial k_y$, which represents the electron's velocity in the y-direction. Since there is no mirror symmetry in the y-direction, the electron velocity shows opposite signs but different magnitudes in the $k_y<0$ and $k_y>0$ regions. As a result, $\beta^y_{xy}({\bf k})$ at $k_y<0$ and $k_y>0$ have opposite signs and do not entirely offset each other. When applying an out-of-plane electric field with gate voltage $V_g=18$V \cite{footnote}, the dipole distribution of the CPGE tensor $\beta^{y,\tau_z}_{xy}({\bf k})$ at $K/K^{\prime}$ valleys reverses, as Fig. \ref{Fig2} (b) shows. In Fig. \ref{Fig2} (c), we present the valley-dependent CPGE tensor $\beta^{y,\tau_z}_{xy}$ vs the gate voltage. As the gate voltage increases, the CPGE tensor $\beta^{y,\tau_z}_{xy}$ of the $K$ valley switches from positive to negative, while the $K^{\prime}$ valley's component changes from negative to positive. This suggests a valley photocurrent switch for certain $\sigma_{\pm}$ circularly polarized light.

Since the CPGE involves $\sigma_{+}$ and $\sigma_{-}$ together, we then calculate the valley-polarized photoresponse at $K$ and $K^{\prime}$ valleys under certain $\sigma_-$/$\sigma_+$ circularly polarized light using the non-equilibrium Green's function combined with density functional theory (NEGF-DFT) formalism\cite{taylor2001,lei2014}. A two-probe transport device is built based on monolayer BiAsI$_2$ valley materials as shown in Fig. \ref{Fig3} (a). In the numerical simulations, the device can be divided into three regions: the central scattering region where the light is shining, and the left and right leads that extend to the electron reservoirs in infinity as shown in Fig. \ref{Fig3} (e). There are five unit cells with a total of 40 atoms in the central scattering region, about 44.23 {\AA}. During the simulation, along the $z$ direction, vacuum layers of 9.5 {\AA} were added to the top and bottom of the device to avoid the fake interaction between neighboring slabs. Here, back gates are applied throughout the system as shown in Fig. \ref{Fig3} (e) to achieve the gate effect by setting a gate-induced electrostatic boundary condition for the Hartree potential when solving the Poisson equation during the self-consistent calculations.


After obtaining the NEGF-DFT self-consistent device Hamiltonian, the valley-polarized photocurrent flowing in the left electrode, $J_{L,\tau_z}^{(ph)}$, can be expressed using the following formula,\cite{Chen2012,lei2014}
\begin{equation}
\begin{aligned}
\label{photocurrent}
J^{(ph)}_{L,\tau_z}&=\frac{e}{h}\sum_{{\bf k}\in \tau_z}\int T_{eff}(\epsilon,{\bf k}) d\epsilon\\
&=\frac{ie}{h}\sum_{{\bf k}\in \tau_z}\int {{\rm Tr}\{\Gamma_{L}[G^{<}_{ph}+f_L(\epsilon)(G^{>}_{ph}-G^{<}_{ph})]\}}d\epsilon,
\end{aligned}
\end{equation}
where \textit{L} indicates the left lead; $\Gamma_L=i(\Sigma_L^r-\Sigma_L^a)$ is the linewidth function and $\Sigma^r_L=[\Sigma^a_L]^{\dagger}$ is the retarded self-energy due to the presence of the left lead; $G^{</>}_{ph}=G_0^{r}\Sigma_{ph}^{</>}G_0^{a}$ represents the lesser/greater Green's function including the electron-photon interaction,\cite{henrickson2002} where the $G_0^{r/a}$ is the retarded/advanced Green's functions without photons and $\Sigma_{ph}^{</>}$ is the self-energy due to the presence of the electron-photon interaction; $f_L(E)$ is the Fermi-Dirac distribution function of the left lead. $T_{eff}(\epsilon,\bf k)$ is the effective transmission coefficient. The polarization of the light can be defined by a complex vector \textbf{e}. For circularly polarized light, $\bf{e}=\frac{1}{\sqrt2}(\bf{e}_1\pm\textit{i}\bf{e}_2)$.\cite{xie2015} In our numerical calculations, the vectors $\bf{e_1}$ and $\bf{e_2}$ are set along the $y$ and $x$ directions and the light is incident along the $-z$ direction as shown in Fig. \ref{Fig3} (e).

For simplicity, we introduce a normalized valley related photoresponse\cite{henrickson2002,Chen2012,chen2018},
\begin{equation}
\label{R_s}
R^{\tau_z}=\frac{J^{(ph)}_{L,\tau_z}}{eI_\omega},
\end{equation}
where the unit of $R^{\tau_z}$ is ${a_0^2}/{photon}$ and $a_0$ represents the Bohr radius; $I_\omega$ is the photon flux defined as the number of photons per unit time per unit area.

In order to characterize the polarization of the generated valley-polarized photocurrent, the valley polarization (VP) is defined as
\begin{equation}
\label{VP}
P_{valley}(\%)=\frac{|R^{K^{\prime}}|-|R^{K}|}{|R^{K^{\prime}}|+|R^{K}|}
\times 100.
\end{equation}

The valley-polarized photoresponse of a monolayer BiAsI$_2$ dependence on gate voltage is shown in Figs. \ref{Fig3} (b, f). We note that without an external out-of-plane electric field, the photoresponse at the $K^{\prime}$ valley can only be stimulated by the $\sigma_-$ circularly polarized light, creating a valley polarization of $100\%$. Yet, when a vertical voltage $V_g$ is introduced, the photoresponse at the $K^{\prime}$ valley decreases, while the photoresponse at the $K$ valley rises. At a gate voltage of $18$V, the photoresponse at the $K^{\prime}$ valley is nearly zero, and the corresponding valley polarization reaches $-100\%$. In contrast, with $\sigma_+$ circularly polarized light, the photoresponse at the $K^{\prime}$ valley grows with the gate voltage, while the photoresponse at the $K$ valley shrinks. Thus, the valley polarization changes from $-100$ to $100\%$ when the gate voltage of $18$V is applied. Figs. \ref{Fig3} (c, g) displays the effective transmission coefficient $T_{eff}$ at the $K^{\prime}$ and $K$ valleys without gate voltage. It is observed that only the electrons at the $K^{\prime}$ valley can transport under the $\sigma_-$ circularly polarized light, whereas the electrons at the $K$ valley transport under the $\sigma_+$ circularly polarized light. However, when the gate voltage $V_g=18$V is used, the electrons at the $K^{\prime}$ valley can transport under the $\sigma_+$ circularly polarized light, while the electrons at the $K$ valley operate under the $\sigma_-$ circularly polarized light, as shown in Figs. \ref{Fig3} (d, h). These directly present a fully valley-polarized photocurrent switch, achieved by applying the out-of-plane electric field to the monolayer BiAsI$_2$.

\begin{figure}
\centering
\includegraphics[scale=0.73]{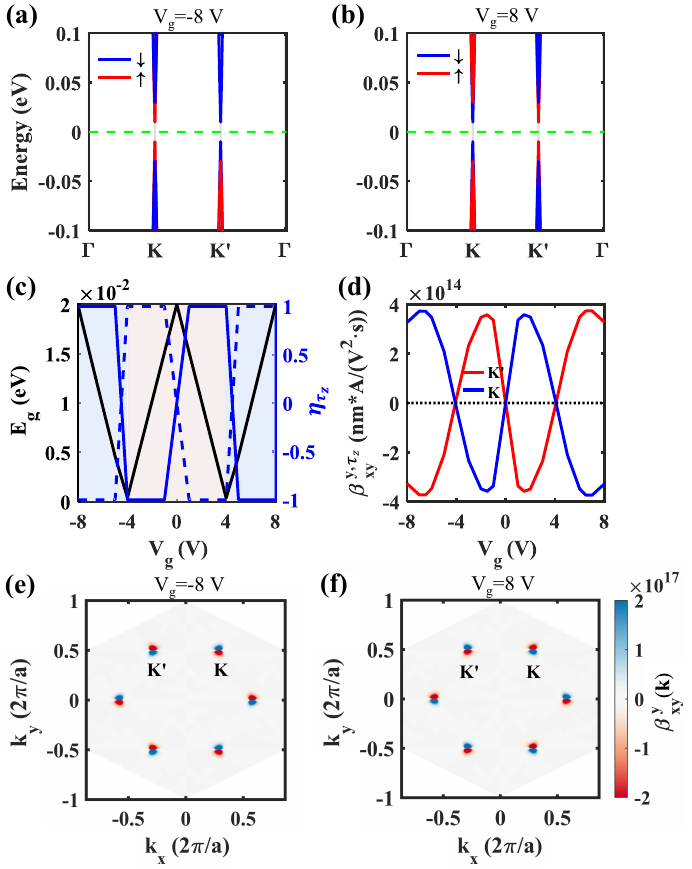}
\caption{(a, b) The band structure of germanene when $V_g=\mp 8$V, respectively. The horizontal green dashed line represents the Fermi energy. (c) The band gap $E_g$ and the degree of circular polarization $\eta_{\tau_z}$ at $K$ and $K^{\prime}$ valleys vs the gate voltage $V_g$. The solid and dashed blue lines represent $K$ and $K^{\prime}$ valleys, respectively. (d) The valley-dependent CPGE tensor $\beta^{y,\tau_z}_{xy}$ vs the gate voltage $V_g$. Here, the incident photon energy is fixed at $\hbar\omega=0.025$ eV. (e, f) The $\bf k$-specified contribution to the CPGE tensor $\beta^{y}_{xy}({\bf k})$ when $V_g=\mp 8$V, respectively.}
\label{Fig4}
\end{figure}

Recently, the TPT induced by the out-of-plane electric field in germanene has been successfully confirmed in experiments.\cite{Arka2022,li2014b,deng2018,Bampoulis2023} Consequently, it is promising to realize the fully valley-polarized photocurrent switch based on germanene. Germanene has inversion symmetry, which means the Berry curvature equals zero. When the out-of-plane electric field is applied, this symmetry is broken, leading to a non-zero Berry curvature. The direction of the electric field can control the sign of the Berry curvature. Without an external electric field, germanene is a direct bandgap semiconductor with a bandgap of $20$ meV. When a negative gate voltage $V_g=-8$V is applied, the maximum valence band at the $K$ valley is spin up, while the $K^{\prime}$ valley is spin down. Reversing the gate voltage $V_g$ to $8$V also reverses the spin of the maximum valence band at the $K/K^{\prime}$ valley, with a TPT transition between QVHE as discussed in the Dirac model analysis, as shown in Figs. \ref{Fig4} (a, b). When the gate voltage changes from $-$8 to 8V, the band gap closes twice, which is accompanied by a reversal of the circular polarization degree's sign, as illustrated in Fig. \ref{Fig4} (c) and Supplemental Materials Fig. S3\cite{supple}. This results in the flipping of the CPGE tensor's sign, which offers the potential to switch the valley photocurrent. The CPGE tensor $\beta^y_{xy}({\bf k})$ at $V_g=\mp 8$V at $K/K'$ valleys displays an opposite dipole distribution, as shown in Figs. \ref{Fig4} (e) and (f). The corresponding dipole distributions at $K/K^{\prime}$ valleys also reverse through the tuning gate voltage.

\begin{figure}
\centering
\includegraphics[scale=0.67]{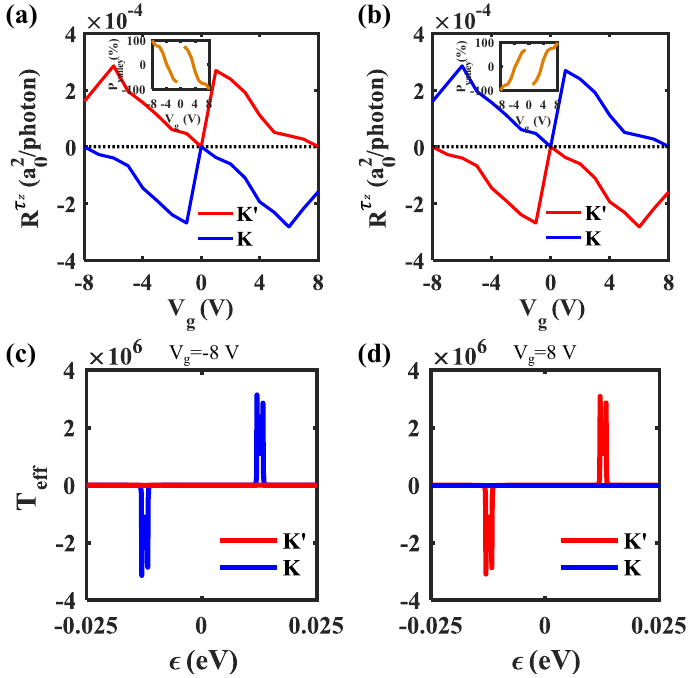}
\caption{The valley-polarized photoresponse $R^{\tau_z}$ vs the gate voltage $V_g$ under the vertically incident (a) $\sigma_-$, (b) $\sigma_+$ circularly polarized light, respectively. Inserts show the corresponding valley polarization vs $V_g$. (c, d) The effective transmission $T_{eff}$ vs $\epsilon$ at $K$ and $K^{\prime}$ points when $V_g=\mp 8$V of the vertically incident $\sigma_+$ circularly polarized light. Here, the incident photon energy is fixed at $\hbar\omega=0.025$ eV.}
\label{Fig5}
\end{figure}

Finally, the valley-polarized photoresponse of germanene dependence on gate voltage is shown in Figs. \ref{Fig5} (a, b). When the gate voltage is zero, the Berry curvatures in the Brillouin zone are zero due to the presence of inversion and time reversal symmetries, which in turn results in a zero photoresponse. However, when an out-of-plane electric field is applied, the broken inversion symmetry leads to a non-zero Berry curvature and subsequently, a non-zero photoresponse. When a gate voltage of $-8$V is applied, the photoresponse at $K^{\prime}$ valley can only be stimulated by the $\sigma_-$ circularly polarized light, creating a valley polarization of $100\%$. Yet, when a vertical voltage $V_g=8$V is introduced, the photoresponse at the $K^{\prime}$ valley is nearly zero, and the corresponding valley polarization reaches $-100\%$. In contrast, with $\sigma_+$ circularly polarized light, the photoresponse at the $K/K^{\prime}$ valley shows the opposite change. Thus, the valley polarization changes from $-100$ to $100\%$ when the gate voltage of $\mp 8$V is applied. When calculating the effective transmission coefficients at $K/K^{\prime}$ valleys with $\sigma_+$ circular polarized light, we find that when $V_g=-8$V, only the electrons at the $K$ valley can transport, while transported electrons at the $K^{\prime}$ valley are forbidden. When $V_g=8$V, the valley-dependent optical selection rule changes. As a result, only the electrons at the $K'$ valley can transport, while the transport of electrons at the $K$ valley are forbidden. This confirms that by controlling the direction of the electric field, we can realize a valley switch based on germanene. It is important to note that the valley-polarized photocurrent produced in germanene is symmetrical with the gate voltage. This is because of the presence of inversion symmetry when there is no gate voltage. In contrast, the valley polarized photocurrent in monolayer BiAsI$_2$ is expected to be asymmetrical due to its lower symmetry.

\emph{Conclusions}-To summarize, we propose a scheme of a valley switch by electric field. This method allows for effective control of valley polarization, resulting in valley photocurrents with opposite valley indices. Using first-principles calculation, we successfully realize the switching of valley-polarized photocurrent using the electric field in monolayer BiAsI$_2$ and germanene. The corresponding valley polarization can be switched between $100$ and $-100\%$ by applying the out-of-plane electric field. This opens up new possibilities for the development of future valley-based electronic storage devices.

$${\bf ACKNOWLEDGMENTS}$$

We gratefully acknowledge the support from the National Key Research and Development Program of China (Grant No. 2022YFA1404003), the National Natural Science Foundation of China (Grants No. 12074230, No. 12034014, and No. 12174231), the Fund for Shanxi ``1331 Project", Fundamental Research Program of Shanxi Province (Grant No. 202103021222001), Research Project Supported by Shanxi Scholarship Council of China, and the Graduate Education Innovation Project of Shanxi Province (Grant No. 2023KY013). This research was partially conducted using the High Performance Computer of Shanxi University.

\bigskip

\noindent{$^{*)}$chenjun@sxu.edu.cn}\\
\noindent{$^{\dagger)}$zhanglei@sxu.edu.cn}\\
\noindent{$^{\ddag)}$jianwang@hku.hk}

\nocite{sipe2000,de2017,xu2018E,Kresse1993,Kresse1996,zhou2021,Arka2022,deng2018,taylor2001,lei2014,nanodcal2,kleinman1982,perdew1996,chen2018,xie2015,Chen2012,henrickson2002,zhang2009,wang2019,weintrub2022,tan2023}

\bibliography{BiAsI2}

\end{document}